\documentclass[usenatbib]{mnras}

\usepackage{multirow}
\usepackage{rotating}
\usepackage{colortbl}
\usepackage{color}
\usepackage[fleqn]{amsmath}
\usepackage{amssymb}
\usepackage{amsfonts}
\usepackage{verbatim}
\usepackage{scalefnt}
\usepackage[percent]{overpic}
\usepackage[T1]{fontenc} 
\usepackage{aecompl} 
\usepackage{ulem}
\usepackage{cleveref}
\usepackage{todonotes}

\usepackage{float}

\def\lsim{\mathrel{\rlap{\lower 3pt \hbox{$\sim$}} \raise 2.0pt \hbox{$<$}}}
\def\gsim{\mathrel{\rlap{\lower 3pt \hbox{$\sim$}} \raise 2.0pt \hbox{$>$}}}

\newcommand{\comments}[1]{} 

\title[Merger Driven Direct Collapse]{Direct collapse of exceptionally heavy black holes in the merger-driven scenario}
\author[Lorenz Zwick et al.]{Lorenz Zwick$^{1}$\thanks{E-mail: zwicklo@ics.uzh.ch}, Lucio Mayer$^{1}$, Lionel Haemmerl\'e$^{2}$, Ralf S.\ Klessen$^{3,4}$
\\
$^{1}$Center for Theoretical Astrophysics and Cosmology, Institute for Computational Science, University of Zurich,\\ 
Winterthurerstrasse 190, CH-8057 Z{\"u}rich, Switzerland\\
$^{2}$D\'epartement d'Astronomie, Universit\'e de Gen\`eve, chemin des Maillettes 51, CH-1290 Versoix, Switzerland\\
$^{3}$Universit\"at Heidelberg, Zentrum f\"ur Astronomie, Intitut f\"ur Theoretische Astrophysik, \\
Albert-Ueberle-Str 2, D-69120 Heidelberg, Germany\\
$^{4}$Universit\"{a}t Heidelberg, Interdisziplin\"{a}res Zentrum f\"{u}r Wissenschaftliches Rechnen,\\
Im Neuenheimer Feld 205, D-69120 Heidelberg, Germany\\
}
\date{Accepted XXX. Received YYY; in original form ZZZ}
\pubyear{2022}
\begin{document}
\label{firstpage}
\pagerange{\pageref{firstpage}--\pageref{lastpage}}
\maketitle


\begin{abstract}
We revisit the conditions present in supermassive discs (SMDs) formed by the merger of gas-rich, metal-enriched galaxies at red-shift $z\sim 10$. We find that SMDs naturally form hydrostatic cores which go through a rapidly accreting supermassive star phase, before directly collapsing into massive black holes via the general relativistic instability. The growth and collapse of the cores occurs within $\sim 5\times 10^5$ yr from the formation of the SMD, producing bright electromagnetic, neutrino and gravitational wave transients with a typical duration of a few minutes and, respectively, a typical flux and a typical strain amplitude at Earth of $\sim 10^{-8}$ erg s$^{-1}$ cm$^{-2}$ and $\sim4\times 10^{-21}$. We provide a simple fitting formula for the the resulting black hole masses, which range from a few $10^6$ M$_{\odot}$ to $10^8$ M$_{\odot}$ depending on the initial SMD configuration. Crucially, our analysis does not require any specific assumption on the thermal properties of the gas, nor on the angular momentum loss mechanisms within the SMD.
Led by these findings, we argue that the merger-driven scenario provides a robust pathway for the rapid formation of supermassive black holes at $z > 6$. It provides an explanation for the origin of the brightest and oldest quasars without the need of a sustained growth phase from
a much smaller seed. Its smoking gun signatures can be tested directly via multi-messenger observations.

\end{abstract}

\begin{keywords}
Hydrodynamics; quasars: supermassive black holes; galaxies: high-redshift;  X-rays: bursts; gravitational waves; methods: analytical;
\end{keywords}


\section{Introduction}
After the first radio observations and key theoretical insights of the 1960ies \citep[see, e.g.][]{Bolton63,Schmidt63,1964zeldovich}, recent surveys have found several quasars with estimated masses exceeding $10^{9}$~M$_{\sun}$ at redshift $z \gsim 7$, i.e. only 700 million years after the Big Bang \citep[see, e.g.][]{2017Mazzucchelli,2020sloan,wang2021}. These observations are both impressive and unexpected. They call for satisfying theoretical models, able to both explain the existence of such behemoths, as well as the uncountable number of smaller supermassive black holes (SMBHs) that likely exist at high redshift. Or perhaps it would be more reasonable for such extreme outliers to form via completely independent mechanisms from the large majority of SMBHs? We will return to this question in the conclusion of this work.

Barring primordial BHs scenarios \citep[for some recent constraints, see, e.g.][]{2018Bellomo,Zhou2022,2022zhou2}, proposed mechanisms for early SMBH formation can be broadly separated into two categories. In the so-called ``light seeds'' scenario, BHs can form at $z\gsim 15$ as the last stage in the evolution of population III stars \citep[see, e.g.][]{Madaurees2001,Abel2002,Schneider2002,Hirano2014}. Despite the large stellar masses, population III stellar evolution models consistently predict compact remains of around $\lsim 10^2$ M$_{\sun}$. Considering the fact that Eddington-limited accretion can typically only double BH masses on a Salpeter time of $\sim 5\times 10^7$~yr \citep[][]{1921eddington,1964Salpeter}, it has become doubtful whether light seed models can satisfyingly explain the enormous masses of high-redshift quasars \citep[see, e.g.][]{2008alvarez,2018smith,Zhu2020,Sassano2021}. A more promising origin for SMBHs is provided by ``direct collapse'' models (DCMs). What defines a DCM is the possibility for large amounts of gas to collapse into a ``heavy seed'', typically with a mass of around $\sim 10^4$ M$_{\rm \sun}$ to $\sim 10^5$ M$_{\rm \sun}$, which provides a significant jump-start for the growth towards higher masses. However, DCMs often require a very specific balance between angular momentum loss and cooling mechanisms in order for gas structures to contract without fragmenting or reaching new equilibrium states. The question of BH growth is also not completely resolved, since several hundred million year stretches of Eddington-limited accretion are still necessary for a $10^5$ M$_{\sun}$ seed to reach observed SMBH masses by $z \sim 7$. While possible in principle, such sustained growth rates require rather compliant gas reservoirs and have (arguably) proven difficult to achieve in simulations where accretion is not assumed a priori. Several works have shown that seeds are only able to grow by a few orders of magnitude before further accretion is progressively quenched by stellar feedback, supernovae or even out-flowing winds \citep[see, e.g.][]{2016MNRASfiacconi,2016lupi,Regan2019,2022sassano}. Due to the many uncertainties in the modelling of high-redshift galaxies, the degree to which some (or all) DCMs might be too idealised is still unclear, and no consensus has yet been reached with regards to which might be the most realistic. We refer the reader to the many insightful reviews on the advantages and disadvantages of different DCMs \citep[see, e.g.][]{2011priya,2012Volonteri,2013Haiman,2019Mayer,Woods2019} and we will mention specific ones throughout this work.

In recent years, Mayer et al. have proposed a scenario in which many of the typical problems specific to DCMs might be largely by-passed. This pathway to the formation of SMBHs originates from simulations of major gas rich galaxy mergers at $z \sim 10$ \citep[][]{mayer2010,2014bonoli,Mayer2015} and is therefore referred to as the ``merger-driven'' DCM. In these simulations, tidal torques and hydrodynamical instabilites caused by the merger induce gas inflow rates in excess of $\sim 10^4$~M$_{\sun}$~yr$^{-1}$. Enormous quantities of gas are rushed into the centre of the newly merged proto-galaxy, where they form a rotationally supported disc-like gas structure. These ``supermassive discs'' (SMDs) typically contain upwards of 10$^9$ M$_{\sun}$, localised in a region of only $\lsim 1$ pc. In contrast to typical proto-galactic discs, they are opaque to radiation, stable to fragmentation and contain up to Solar metallicity due to their relatively late formation. The argument in \cite{mayer2010} and \cite{Mayer2015} (hereafter MI and MII, respectively) is that SMDs represent an optimal environment for the rapid formation of an SMBH, either through an intermediate supermassive star (SMS) phase \citep[see, e.g.][]{Lionel2020,lionel2021}, or directly via the collapse of a general relativistic (GR)-unstable region. The advantages of the merger-driven scenario are that the formation of SMDs does not require any specific restrictions on thermodynamical conditions of the gas, nor any specific angular momentum loss mechanism other than the ones provided by the hydrodynamics of galaxy mergers. However, the exact process by which an SMD might produce a BH is unspecified, since its formation is likely to occur well below the resolution limit of the aforementioned simulations. The aim of this work is thus set: revisit the peculiar conditions present in SMDs to determine the mass of BHs predicted by the merger-driven scenario.\newline \newline

The paper is structured as follows: In Section~\ref{sec:TB}, we briefly discuss the peculiar initial conditions, time-scales and physical processes that govern the evolution of SMDs. In Section~\ref{sec:model}, we use thermodynamical arguments to describe the accretion of the SMD onto a hydrostatic core via a generic angular momentum loss mechanism. We derive a mass-radius-temperature relation for the core and show that it can grow to significant size before approaching the GR instability regime. We provide a simple fitting formula for the mass of the resulting BHs, which have a typical scale of a few 10$^7$ M$_{\sun}$. In Section~\ref{sec:discussion}, we discuss the implications as well as some caveats of our model. We focus on the direct observational consequences of the merger-driven DCM, highlighting possible electromagnetic and gravitational counterparts. Finally, we summarise our findings and present some concluding remarks in Section~\ref{sec:conclusion}.

\begin{figure}
    \centering
    \includegraphics[scale=0.3]{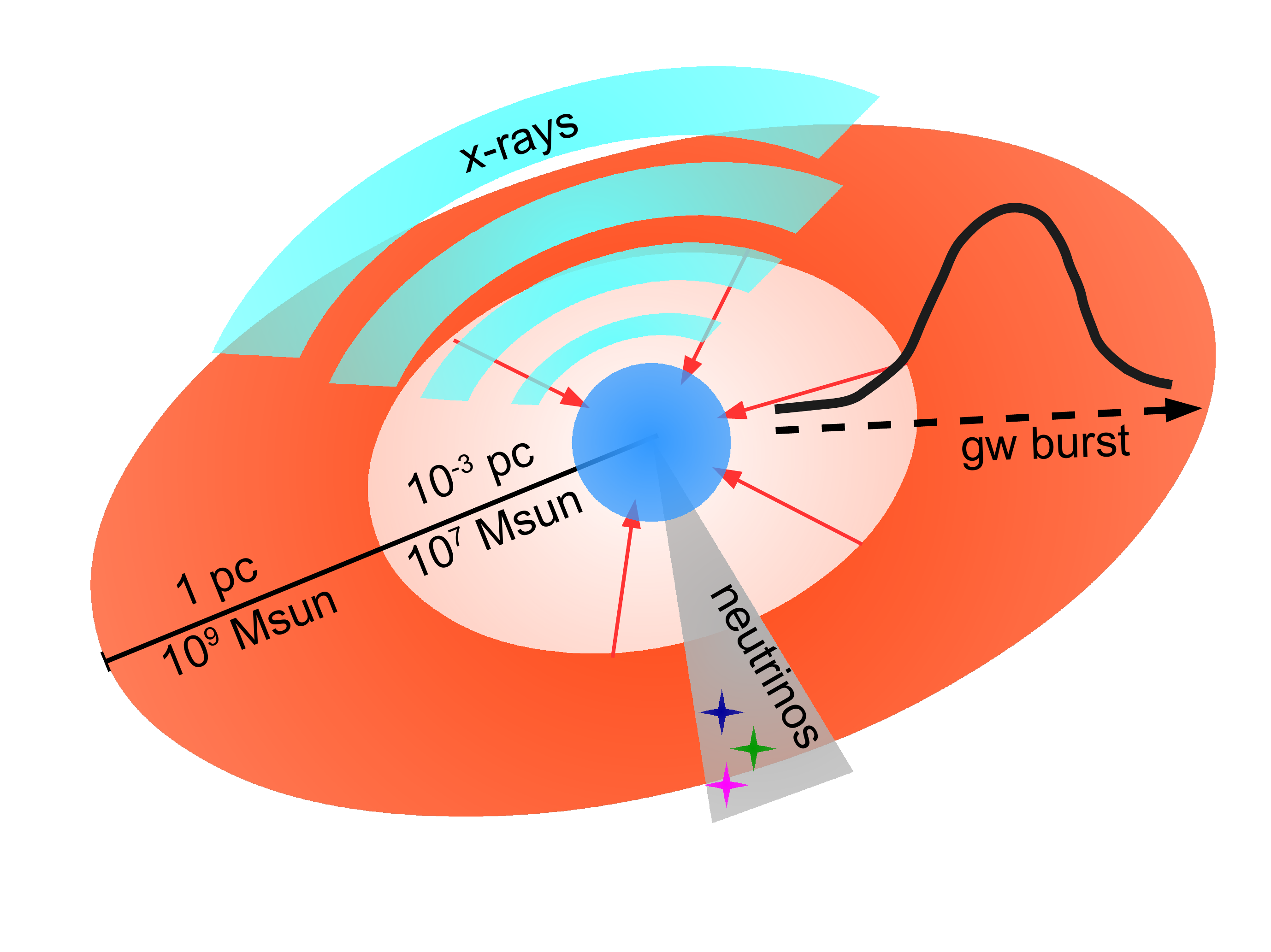}
    \caption{A simple cartoon of the merger-driven DC scenario and its observational signatures. An SMD with a typical size of 1 pc and a typical mass of $10^9$ M$_{\sun}$ loses angular momentum by a secular or dynamical mechanism (see section \ref{sec:TB_angmom}). The central regions accrete onto a hydrostatic core, which increases in size, mass and internal energy. The core reaches the GR instability at a mass of a few $10^7$ M$_{\sun}$ (see section \ref{sec:AM_collapse}). It then rapidly collapses into a fully fledged SMBH, producing a burst of x-rays, gravitational waves and neutrinos along the way (see section \ref{sec:observation}).}
    \label{fig:cartoon}
\end{figure}

\section{The physics of supermassive discs}\label{sec:TB}
\subsection{The initial configuration}\label{sec:TB_supermassive_disc}
The formation of SMDs has been discussed extensively in several works (MI; \citealt{mayer2010}; MII; \citealt{Mayer2015}, \citealt{2014bonoli}, \citealt{2019Mayer}). For the purposes of this work, we assume that the final stages of the aforementioned multi-scale simulations are representative of the fate of gas in major galaxy mergers at $z \sim 10$. In particular, we take the SMDs analysed in MII as a baseline case, but allow for important parameters such as mass, size and temperature to vary slightly from the simulated results, in case the latter are contaminated by numerical artefacts and spurious sub-grid physics.

During the final stages of the aforementioned simulations, large quantities of gas settle into a rotationally supported SMD containing upwards of $10^9$ M$_{\sun}$. The mass is localised in a region of size $\sim 1$ pc, yielding a typical average density of $\rho \sim 10^{-11}$ Kg/m$^3$, or $ \sim 10^{10}$ particles per cubic centimetre. 
The gravitational compactness $\mathcal{K}_{\rm d}$, is defined as the size of an object divided by the Schwarzschild radius \citep[][]{schwarzschild} of its total mass. Typical values for SMDs read:

\begin{align}
    \label{eq:compactness}
    \mathcal{K}_{\rm d} = \frac{c^2 R_{\rm d}}{2 G M_{\rm d}} \sim 10^{4} \left(\frac{10^9 \, \rm{M}_{\sun}}{M_{\rm d}} \right) \left(\frac{R_{\rm d}}{1 \, \rm{pc}} \right),
\end{align}

\noindent where $M_{\rm d}$ is the mass of the SMD and $R_{\rm{d}}$ its radius\footnote{SMDs are as gravitationally compact as the typical white dwarf, but with densities twenty order of magnitudes lower \citep[][]{1925adams,1990koester}.}. The dynamical time-scale $t_{\rm{dyn}} \sim \rho^{-1/2}$ of such a system is close to 500~yr, only two orders of magnitude larger than the light-crossing time $t_{\rm c} \sim R_{\rm d} c^{-1} \sim 3$~yr, where $c$ is the speed of light. The unusual compactness of SMDs is the primary motivation for considering the GR instability criterion, which will be discussed in more detail in Section~\ref{sec:TB:GRinst}. A crucial difference between SMDs and more conventional proto-galactic discs is the fact that they form at the transition between the optically thin and optically thick regimes. Therefore, any further contraction will be strongly affected by adiabatic heating. Consider the mean free path $\lambda_{\rm{ph}}$ of a photon that undergoes Thomson scattering \citep[][]{1851thomson}:

\begin{align}
    \label{eq:mfp}
     \lambda_{\rm{ph}} =  \frac{1}{n_{\rm e} \sigma_{\rm{T}}} \approx \mathcal{A} \times 10^{-5} \left(\frac{10^9 \, \rm{M}_{\sun}}{M_{\rm d}} \right) \left(\frac{R_{\rm d}}{1 \, \rm{pc}} \right)^3 \, \left[ \rm{pc}\right],
\end{align}

\noindent where $\mathcal{A}$ corresponds to the average aspect ratio of the disc (which typically ranges from $\sim 10^{-2}$ to $10^{-1}$), $n_{\rm e}$ is the number density of free electrons, $\sigma_{\rm T}$ is the Thomson cross-section and we assumed full ionisation, which is appropriate for low density hydrogen at these temperatures \citep[][]{1921saha}. Eq.~\ref{eq:mfp} seems to suggest that SMDs should be considered opaque. However, a system with a short mean free path can still lose heat over a sufficiently long period of time. What determines if a system truly behaves adiabatically is the diffusion time-scale $t_{\rm{diff}}$:

\begin{align}
\label{eq:difftimes}
t_{\rm{diff}} = \frac{\mathcal{A}^2R_{\rm d}^2}{\lambda_{\rm{ph}}^2} \frac{\lambda_{\rm{ph}}}{c} \approx 10^5 \mathcal{A} \left(\frac{M_{\rm d}}{10^9 \, \rm{M}_{\sun}} \right) \left(\frac{1 \, \rm{pc}}{R_{\rm d}} \right) \, \left[ \rm{yr}\right],
\end{align}

\noindent where we approximate photon paths as random walks. Due to the enormous accretion rates triggered by the galaxy merger ($\sim 10^4$~M$_{\sun}$~yr$^{-1}$), more and more mass is accumulated within a given central region, enabling the formation of an SMD in approximately $10^5$~yr, a time comparable to Eq.~\ref{eq:difftimes}. This suggests that the assembly of an SMD occurs mostly in an ``effectively'' optically thin regime, where radiation has time to escape the system before much more mass is added. The system finally becomes truly opaque only when a sufficient amount of mass ($\sim 10^9$ M$_{\sun}$) is concentrated within a region of a fixed scale ($\sim 1$ pc). From this point on, the accretion of even more mass would not change the thermodynamic behaviour of the disc qualitatively, since it is already adiabatic as a whole\footnote{The addition of orders of magnitudes more mass might indeed change the thermodynamics of the SMD. This is however not a realistic possibility since the size of dark matter haloes and their gas reservoirs are limited in size at $z \sim 10$.}.
Interestingly, this simple consideration yields a mass-radius relation for SMDs, which sets a natural density scale that nicely aligns results reported in MI (whereas it slightly differs from the results in MII, in which SMDs are more compact by a factor few). 

The initial thermal state of an SMD is determined by the dynamics of its formation. The competition between atomic cooling, shock heating and turbulent dissipation results in a disc which is approximately isothermal at a temperature of $T_{\rm d}\sim 7 \times 10^3$ K, in which fragmentation is suppressed regardless of the presence of metals. Despite the different physics at play, this final state is comparable to the virial temperatures commonly expected in smaller, metal-free proto-galactic discs which are often invoked in DCMs \citep[see, e.g.][]{1977rees,1991white}. At typical temperatures of $\lsim 10^4$ K, the total thermal energy of the gas is not sufficient for spherical hydrostatic equilibrium and the inflows settle into a largely rotationally supported disc. In MI and MII, the resulting SMDs display a constant scale-height of approximately $\sim 0.1 R_{\rm d}$, likely due to the resolution limit. In this paper, we assume the standard scaling for the disc's scale-height, $H \sim c_{\rm s}/\omega$, where $\omega$ is the angular velocity and $c_{\rm s}$ the isothermal speed of sound for a mixture of gas and radiation.

Any self gravitating fluid is likely to exhibit some degree of mass concentration, i.e. a density profile of the form $\rho \sim r^{-n}$. Pressure-less thin discs with arbitrary density profiles have been famously studied in \cite{1963Mestel}, and the allowed rotational configurations range from constant angular velocity ($\rho \sim r^{0}$ and $v_{\rm t} \sim r$) to constant tangential velocity ($\rho \sim r^{-2}$ and $v_{\rm t} \sim r^0$). A steep density profile also implies that the inner parts of SMDs are likely entirely opaque. Accordingly, the diffusion time-scale within a given sub-region of the SMD must be larger relative to the whole system. Matter concentration can also induce substantial variations in the balance between radiation pressure and gas pressure, leading to a more complex behaviour of the sound speed and the scale-height. Temperature fluctuations can potentially also affect the balance of pressure. Whenever a system is optically thick however, radiative heat transfer can be approximated via the Rosseland diffusion equation \citep[][]{rosseland,Mihalas&Mihalas}. One consequence of diffusion is to smooth out any sufficiently strong temperature variability, suggesting that the isothermality seen in MI and MII is also a good approximation at scales that are below the resolution limits.

\subsection{Mechanisms of angular momentum loss}\label{sec:TB_angmom}

In order to contract, a rotationally supported disc must dissipate large amounts of rotational energy and angular momentum. Various scenarios for the disposal of angular momentum have been proposed in the context of DCMs. For the purposes of this paper, we broadly separate them in the two categories of \textit{secular} and \textit{dynamical} angular momentum loss mechanisms. Secular mechanisms involve the slow transformation of rotational energy into internal energy. The contraction of the disc occurs through a succession of quasi-equilibrium states, over a period of several dynamical time-scales. Examples in the literature include friction against the cosmic radiation background \citep[][]{1993umemura} as well as viscous dissipation caused by turbulence \citep[][]{1995eisenstein,Self_similar_disk,1997mineshige}. In many of these scenarios however, cooling and fragmentation occur on much quicker time-scales than what is required for the disc to contract significantly. This adds the requirement of a nearby reservoir of heat which can influence the energy balance of the disc, in particular the ionisation fraction of hydrogen \citep[commonly by a source of Lyman-Werner radiation, see, e.g.][]{agarwal,2013Latif,2014regan,2016habouzit,2016agarwal,2017johnson}. Dynamical mechanisms involve the rapid transport of angular momentum, inducing a contraction of the disc on time-scales comparable with gravitational free-fall. While these processes can potentially circumvent the issue of fragmentation, they are conditional to specific triggers \citep[such as a nearby supernova explosion,][]{Colgate2003} or to the onset of global gravitational instabilities \citep[e.g.][]{2004koushippas,2006begelman,2006priya,2007lodato,2008begelman}. A mechanism based on cold, turbulent gas inflows has been proposed recently and could circumvent some of the aforementioned challenges, producing DC BHs with masses of the order $\sim 10^4$ M$_{\odot}$ \citep[][]{2022latif}.

Crucially, SMDs provide a realistic cosmological object in which not only dynamical, but also secular mechanisms generally act on time-scales comparable to the diffusion time-scale. The ratio between the diffusion time-scale and dynamical time-scales is at least $\sim \mathcal{A} \times 10^3$, and increases dramatically when considering denser sub-regions of the disc. To assess secular mechanisms, we consider the example of viscous dissipation by means of magnetic gravito-turbulence, a key player in the formation of SMDs (as discussed in MI and MII) that is found to be ubiquitously important in astrophysical disc \citep[see, e.g.][]{1995hawley,2007pessah,2013abramowicz,2013prieto}. The ratio of the diffusion and viscous dissipation time-scales $t_{\rm{visc}}$ can be estimated as follows:

\begin{align}
\label{eq:diffvisc}
    \frac{t_{\rm{diff}}}{t_{\rm{visc}}} \sim \alpha \mathcal{A}^2 \frac{\mathcal{A} R_{\rm d}}{\lambda_{\rm{ph}}} \frac{c_{\rm s}}{c},
\end{align}

\noindent where we assumed an $\alpha$-viscosity prescription \citep[][]{1973Shakura} and used the standard scaling $t_{\rm{visc}} \sim  \alpha^{-1}R_{\rm d}^2 c_{\rm s}^{-1} H^{-1}$. At a temperature of a few thousand Kelvin, the speed of sound in gases is generally much smaller than the speed of light. Consider however, the ratio of thermal energy in a typical SMD that is stored in radiation ($E_{\rm{rad}}$) rather than gas ($E_{\rm{gas}}$):

\begin{align}
    \label{eq:radvsgas}
    \frac{E_{\rm{rad}}}{E_{\rm{gas}}} \sim \mathcal{A} \times 10^3\left(\frac{10^9 \, \rm{M}_{\sun}}{M_{\rm d}} \right) \left(\frac{R_{\rm d}}{1 \, \rm{pc}} \right)^3\left(\frac{T_{\rm{d}}}{7000 \, \rm{K}} \right)^3,
\end{align}

\noindent scaled here for values typical to SMDs. Whenever a system is radiation-pressure dominated, the speed of sound approaches the value appropriate for a photon fluid, i.e $c/\sqrt{3}$ \citep[see, e.g.][]{Mihalas&Mihalas}. This implies that the ratio $c_{\rm s} c^{-1}$ in Eq.~\ref{eq:diffvisc} is actually a fraction of order unity rather than a small number. If we assume values for the $\alpha$ parameter of $\sim 0.2$, which are expected in fully ionised discs \citep[see, e.g.][]{2004lodato,2007king}, we can evaluate Eq.~\ref{eq:diffvisc} for a typical SMD:

\begin{align}
    \label{eq:diffviscev}
    \frac{t_{\rm{diff}}}{t_{\rm{visc}}} \sim  \mathcal{A}^2 10^4 \left( \frac{M_{\rm d}}{10^9 \, \rm{M}_{\sun}}\right) \left( \frac{1 \, \rm{pc}}{R_{\rm{d}}}\right)^2,
\end{align}

\noindent where we replaced the number $\alpha c_{\rm s}c^{-1}$ as a fraction of order $10^{-1}$, consistently with the arguments above. Eq.~\ref{eq:diffviscev} shows that, for reasonable values of the aspect ratio, SMDs are expected to dissipate their rotational energy in a time that is comparable or smaller than a diffusion time-scale, i.e. smaller than at most $\sim \mathcal{A}\times 10^5$~yr. Even more so when considering sub-regions of the disc, where the viscous time-scale decreases substantially and the diffusion time-scale is affected by the higher average density. To re-iterate, the dynamical time-scale in SMDs is typically of the order 500~yr, while both the diffusion and the viscous time-scales are on the order of $5\times 10^5$~yr. The crucial implication of Eq.~\ref{eq:diffviscev} is that the merger-driven scenario does not require a dynamical angular momentum loss mechanism, nor a source of ionising radiation in order to circumvent the problem of fragmentation. Rather, the evolution of the SMD naturally occurs in a regime where cooling is negligible, i.e. the adiabatic regime, even in the slowest possible case of secular angular momentum loss.

\subsection{GR instability}\label{sec:TB:GRinst}

The unlikely combination of a high gravitational compactness and a low density is suggestive that SMDs should be close to the GR-unstable regime (without considering the effect of their geometry). The GR instability relies on the relativistic corrections to the equation of radial momentum, that translate into an increase of the critical value for the first adiabatic exponent $\Gamma_1$, above the Newtonian value of 4/3. An analytic criterion for GR instability has been derived by \cite{1964chandra} for the general case of a hydrostatic, spherical mass.
This increase scales with gravitational compactness of the system, For a polytrope (i.e. for homogeneous entropy distribution) the instability is reached when:

\begin{align}
    \Gamma_1 - \frac{4}{3} <  \kappa \frac{2 G M}{c^2 R},
\end{align}

\noindent where $M$ is the system's mass, $R$ its radius and $\kappa$ is a factor of order unity that depends on the polytropic index.
The GR instability is global, i.e it is triggered at the scale of the whole system. The expectation is therefore that the collapse cannot be halted until a compact object is formed. Several numerical GR simulations have confirmed this, showing that a large fraction of the cloud's mass (60\% to 90\%) is retained in the final BH \citep[see, e.g.][]{2009saijo,2013reisswig}. In the context of DCMs, the GR instability is invoked as a mechanism to destabilise supermassive stars, which can potentially lead to the formation of highly massive BHs \citep[see, e.g.][]{2020lionel,2021lionel,2021woods}. The adiabatic index of a mixture of gas and radiation \footnote{The description of a mixture of gas and radiation technically requires three separate adiabatic indexes. Here we only discuss the relevant one for the GR instability criterion.} depends on its temperature and density. It is determined by the ratio between gas pressure and the total pressure, $\beta$, which is approximated by the following expression in the case of radiation dominated gases:

\begin{align}
    \label{eq:adiabaticindex}
    \Gamma_1 = \frac{32 - 27\beta}{24 -21\beta} \approx {4\over3}+{\beta\over6}.
\end{align}

\section{The fate of supermassive discs}\label{sec:model}
\subsection{Hydrostatic cores in SMDs}\label{sec:AM:core_env}

\begin{figure}
    \centering
    \includegraphics[scale=0.7]{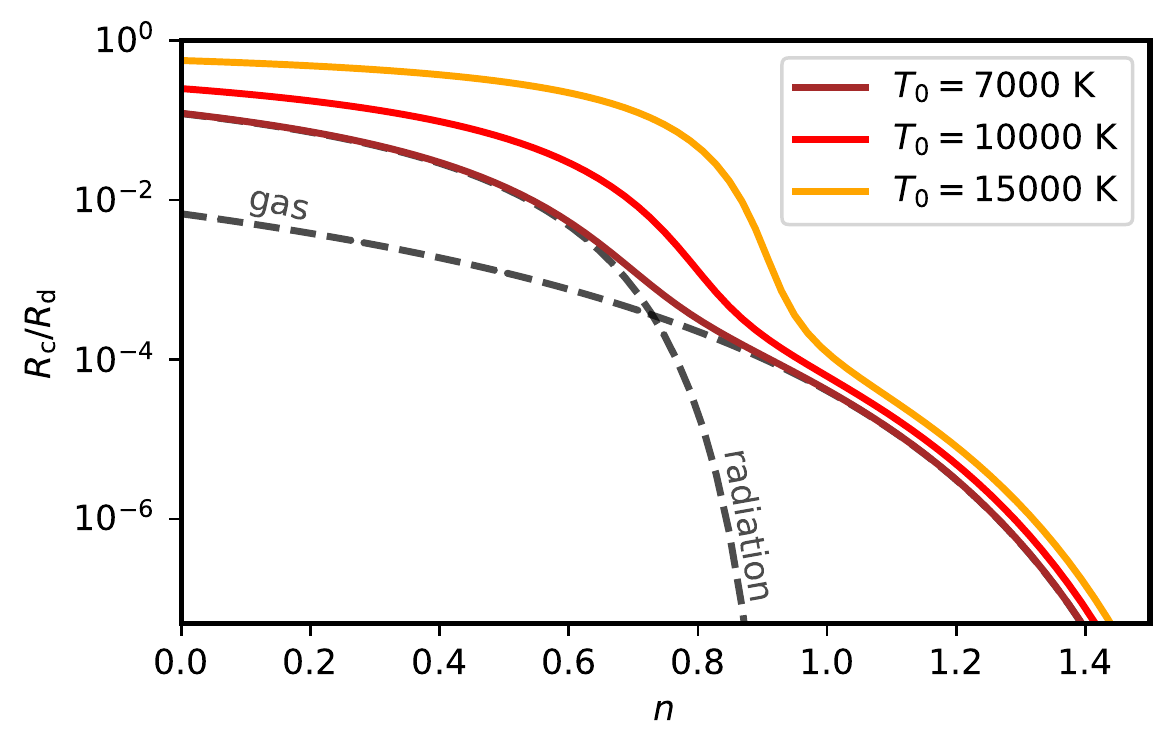}
    \caption{The approximate size of the initial hydrostatic core in an SMD with an initial mass of $10^9$ M$_{\sun}$ and a radius of 1 pc. The numerical results are plotted as coloured lines as a function of the mass concentration parameter $n$(see Eq.~\ref{eq:Mencl}). Clearly visible is the transition betewen gas pressure and radiation pressure dominated cores for a value of $n\sim 0.8$. The grey dashed lines denote the analytical solutions Eq.~\ref{eq:coregas} and Eq.~\ref{eq:corerad} (shown for a temperature of 7000 K).}
    \label{fig:init_core_radius}
\end{figure}

While SMDs are initially supported by rotation, we expect pressure to play a larger and larger role as rotational energy is lost or dissipated. We model the total pressure $P$ within the fluid as a mixture of gas ($P_{\rm{gas}}$) and radiation ($P_{\rm{rad}}$) terms, as is appropriate for non-relativistic radiative flows \citep[][]{Mihalas&Mihalas}:

\begin{align}
    \label{eq:Ptot}
    P &= P_{\rm{gas}}+P_{\rm{rad}},\\
    \label{eq:gas_pressure}
    P_{\rm{gas}}&= 2 \frac{ \rho k_{\rm B} T}{m_{\rm p}},\\
    \label{eq:radiation_pressure}
    P_{\rm{rad}}&=\frac{4 \sigma_{\rm B}}{3 c}T^4.
\end{align}

\noindent Here $m_{\rm p}$ is the mass of a proton, $k_{\rm{B}}$ is the Boltzmann constant, $\sigma_{\rm B}$ is the Stefan-Boltzmann constant, $T$ is the temperature, $\rho$ is the density and the factor 2 in the gas pressure accounts for both electrons and protons\footnote{Electrons and protons contribute equally to gas pressure in an ideal, relaxed plasma \citep[see, e.g.][]{Stolzmann2001}}. The internal specific energy per unit mass, $\mathcal{E}$, is split into a gas and a radiation contribution:

\begin{align}
    \label{eq:internal_energy}
    \mathcal{E} &=\mathcal{E}_{\rm{gas}}+\mathcal{E}_{\rm{rad}}\\
    \label{eq:gas_energy}
    \mathcal{E}_{\rm{gas}}&= \frac{3 P_{\rm{gas}}}{2 \rho},\\
    \label{eq:radiation_energy}
    \mathcal{E}_{\rm{rad}}&=\frac{3 P_{\rm{rad}}}{\rho}.
\end{align}

\noindent The relative dominance between gas and radiation is determined by the temperature and density of every fluid element. We model the latter by defining an enclosed mass profile $M_{\rm r}$ for the SMD, which we approximate as a simple power law:

\begin{align}
    \label{eq:Mencl}
    M_{\rm r} = M_{\rm d} \left(\frac{r}{R_{\rm d}} \right)^{3-n}.
\end{align}

\noindent Here the radius $r < R_{\rm d}$ is an Eulerian radial coordinate defined within the interior of the SMD. The parameter $n$ determines the degree of mass concentration, roughly corresponding to a density profile with a scaling of $\rho \sim r^{-n}$, or a tangential velocity profile of $v_{\rm t} \sim r^{1 - n/2}$. Since all values of $n$ between 0 and 2 are consistent with self-gravitating, rotationally supported discs, we keep $n$ as a free parameter. Higher resolution simulations of the formation of SMDs can reveal if specific values of $n$ are preferred and our results can be evaluated accordingly.

Given Eqs.~\ref{eq:Ptot} to~\ref{eq:Mencl}, we can find a characteristic radius at which the enclosed thermal energy $E_{\rm T}$ is equal to the enclosed gravitational energy $E_{\rm G}$. This is an important boundary which separates the region of an SMD that is rotationally supported from the one that is supported by gas or radiation pressure. By definition, the structure of the latter should not be strongly affected by rotation and can be described as an approximately spherical core in hydrostatic equilibrium. We estimate the core radius $R_{\rm c}$ by setting $E_{\rm G} \sim E_{\rm T}$, and solving for the radius:

\begin{align}
    \label{eq:coreeq}
    E_{\rm G} \sim \int \frac{G M_{\rm{r}} \rho_{\rm{d}}}{r} dV \sim E_{\rm T} \sim \int P dV.
\end{align}

\noindent Here $G$ is Newton's constant and $\rho_d$ is a density profile consistent with Eq.~\ref{eq:Mencl} and the scale height of the SMD. Simple explicit solutions for the core radius can be found by considering either gas or radiation pressure:

\begin{align}
    \label{eq:coregas}
    &\left( \frac{R_{\rm{c}}^{\rm{gas}}}{R_{\rm{d}}}\right)^{2-n} =2\frac{5-2n}{3-n} \frac{k_{\rm B}T}{GMm_{\rm{P}}}R_{\rm d} \\ &\approx 2.7 \times 10^{-5} \frac{5-2n}{3-n} \left(\frac{10^9 \, \rm{M}_{\sun}}{M} \right) \left( \frac{R_{\rm d}}{\rm{pc}} \right)\left( \frac{T}{7000 \, \rm{K}} \right) \nonumber
    \\
    \label{eq:corerad}
    &\left( \frac{R_{\rm{c}}^{\rm{rad}}}{R_{\rm{d}}}\right)^{2-2n} =\frac{16\pi}{9} \frac{5-2n}{3-n} \frac{\sigma_{\rm B} T^4}{c G M^2}R_{\rm d}^4\\
    &\approx 8.8 \times 10^{-3} \frac{5-2n}{3-n} \left(\frac{10^9 \, \rm{M}_{\sun}}{M} \right)^2 \left( \frac{R_{\rm d}}{\rm{pc}} \right)^4\left( \frac{T}{7000 \, \rm{K}} \right)^4, \nonumber
\end{align}

\noindent while the general solution can be found by solving Eq.~\ref{eq:coreeq} numerically.  Eqs.~\ref{eq:coregas} and~\ref{eq:corerad} should be understood as describing the mass and size of a virialised region, which will likely settle into a hydro-static core resembling a radiation dominated supermassive star (SMS) on a local dynamical timescale. Within the scope of this work, we will not be able to precisely model the evolution and the internal structure of these virialised regions, and will make no distinction between them and the SMS they might produce. We discuss this caveat in detail in section~\ref{sec:caveats}, and  expect our methods to correctly capture the relevant global parameters such as total mass, energy content and approximate size.

The results of Eq.~\ref{eq:energypath} are visualised in Fig.~\ref{fig:init_core_radius} for several values of the initial temperature of the SMD. The size of the hydrostatic core varies significantly depending on the parameter $n$, i.e. the degree of matter concentration: SMDs with shallower density profiles ($n \lsim 1$) form large radiation pressure dominated cores, while more concentrated discs form smaller, gas pressure dominated cores. The presence of such a core suggests that the inevitable contraction of SMDs caused by angular momentum loss should be thought as an accretion process onto a spherical, pressure supported central region.

\subsection{Accretion of the disc onto the core}\label{sec:AM:accretion}

In order to properly describe the accretion of mass and energy onto the central core, we consider a shell of fluid elements that is loosing or dissipating rotational energy by the mechanisms detailed in Section~\ref{sec:TB_angmom}. In order for a shell to contract radially, an amount of specific rotational energy proportional to the enclosed gravitational energy must be removed from every fluid element. Because of the arguments detailed in Section~\ref{sec:TB_angmom}, we expect sub-regions of the SMD to contract adiabatically, i.e. for cooling losses to be neglectable. Furthermore, in the case of secular angular momentum loss mechanisms, rotational energy is slowly dissipated within the gas, adding a heating term to its energy balance. We can model all contributions to the internal energy with the following equation:

\begin{align}
    \label{eq:energypath}
    \mathcal{E}(x) = \mathcal{E}_{\rm{gas}} \left(\frac{r_0}{x} \right)^{2}+ \mathcal{E}_{\rm{rad}} \left(\frac{r_0}{x} \right) + \int_{x}^{r_{0}} \frac{G M_{\rm r}}{r^2} \, dr,
\end{align}

\noindent where $r_0$ is the initial radius and $x$ is the current radius of the fluid element. Here the first two terms describe the adiabatic heating of gas and radiation specific energy from the initial values $\mathcal{E}_{\rm{gas}}$ and $\mathcal{E}_{\rm{rad}}$\footnote{Technically, the adiabatic gas and radiation energy components do not increase independently of each other in a thermally coupled gas. This approximation does not however affect our results, since it is the total internal energy that determines the hydrostatic equilibrium configuration.}, while the integral term accounts for the dissipated rotational energy. The latter must be proportional to the gravitational energy at the radius $r$ in order for rotational support to be significantly affected.

In the case of dynamical mechanisms, which can \textit{transport} angular momentum without necessarily dissipating its associated energy, we still expect the integral term in Eq.~\ref{eq:energypath} to appear alongside the adiabatic contributions.
The reason for this is slightly subtler and is due to an effect observed in contracting, self-gravitating fluids known as  turbulent support \citep[see, e.g.][]{1981larson,2015murray,GuangXing18,xu2020}. If a seed of turbulence is present within a fluid element (which is the case in SMDs), any global contraction causes it to heat up, i.e to increase in turbulent mean-root-square velocity \citep[][]{Robertson&Goldreich_2012,2021hennebelle}. In the case of radial self-gravitating flows, turbulence is thus driven to a steady state in which a significant fraction of the inflows' kinetic energy is redirected into turbulent eddies, supporting the system against its own gravity. Since the inflow velocity is ultimately sourced by the gravitational potential, the amount energy that is dissipated by turbulent viscosity will necessarily be proportional to the gravitational energy. Essentially, turbulent heating acts to ``locally virialise'' the internal energy in a contracting fluid, leading to an integral term in Eq.~\ref{eq:energypath} even in the case of a dynamical angular momentum loss mechanism (with a possible prefactor of order unity, which we neglect). The interesting implication of this local virialisation is that the energy content of fluid elements is unaffected by the details of the angular momentum loss mechanism, in particular the time-scale on which it acts (provided it is shorter than a diffusion time-scale, as discussed in Section~\ref{sec:TB_angmom}). This is a stark contrast between the merger-driven scenario and other DCMs and is partially responsible for the robustness of the predicted BH masses in Section~\ref{sec:AM_collapse}.

Eq.~\ref{eq:energypath} shows that, as a gas shell contracts radially, it will accrue a certain amount of internal energy along its path. This contraction is halted only once the conditions for a new equilibrium configuration are reached, i.e when the total internal energy within a certain radius is comparable to the gravitational self energy.
For the considered density profiles,
this condition can be approximated mathematically in terms of the specific energy, which reads:

\begin{align}
    \label{eq:energy_contraction}
    \mathcal{E} (x) \sim \frac{G M_{x}}{x}.
\end{align}

\noindent We can solve Eq.~\ref{eq:energy_contraction}, in combination with Eq.\ref{eq:energypath}, to find the new equilibrium radius of every shell of fluid elements within the SMD. If the system is dominated by gas pressure, the new equilibrium radius $r_{\rm{eq}}$ reads:

\begin{align}
    \label{eq:contraction_gas}
    r_{\rm{eq}}^{\rm{gas}} = r_0\left( \frac{2 k_{\rm B} R_{\rm d} T_0}{G M m_{\rm P}} \right)^{1/2} \left( \frac{R_{\rm{d}}}{r_{\rm s}}\right)^{(2-n)/2}.
\end{align}

\noindent If it is radiation pressure dominated, it reads:

\begin{align}
    \label{eq:contraction_rad}
    r_{\rm{eq}}^{\rm{rad}} = r_0\left( \frac{8 \pi  R_{\rm d}^4 \sigma_{\rm B} T_0^4}{3 c G M^2(3-n)} \right) \left( \frac{R_{\rm{d}}}{r_{\rm s}}\right)^{(2-2n)}.
\end{align}

\noindent The general solution can be found by solving Eq~\ref{eq:energy_contraction} numerically.
Eq.~\ref{eq:contraction_gas} and Eq.~\ref{eq:contraction_rad} are describing how SMDs transition from rotationally supported discs to smaller, spherical gas clouds in hydrostatic equilibrium. This is exactly what is expected, since SMDs lose angular momentum while retaining most of their thermal energy (see Eq.~\ref{eq:difftimes}). The picture is thus one of a growing core embedded within a gaseous disc that contracts from the inside outwards. We define a contraction factor, $\mathcal{K}$, which encodes the global change in geometry from a disc of radius $R_{\rm d}$ to a spherical cloud of radius $\mathcal{K} R_{\rm d}$ and is found by evaluating either Eq.~\ref{eq:contraction_gas} or Eq.~\ref{eq:contraction_rad} at the disc radius $R_{\rm d}$. An explicit expression can be calculated for a gas pressure or radiation pressure dominated system:

\begin{align}
    \label{eq:contractionfactorgas}
     &\mathcal{K}_{\rm{gas}} = \left( \frac{2 k_{\rm B} R_{\rm d} T_0}{G M m_{\rm P}} \right)^{1/2} \\ &\approx 5.2 \times 10^{-3} \left(\frac{R_{\rm d}}{\rm{pc}} \right)^{1/2} \left(\frac{10^9 \, \rm{M}_{\sun}}{M} \right)^{1/2} \left(\frac{T}{7000 \, \rm{K}} \right)^{1/2}, \\
     \label{eq:contractionfactorrad}
     & \mathcal{K}_{\rm{rad}}= \left( \frac{8 \pi  R_{\rm d}^4 \sigma_{\rm B} T_0^4}{3 c G M^2(3-n)} \right)
     \\ & \approx 1.3 \times 10^{-2} \left(\frac{R_{\rm d}}{\rm{pc}} \right)^4 \left(\frac{10^9 \, \rm{M}_{\sun}}{M} \right)^{2} \left(\frac{T_0}{7000 \, \rm{K}} \right)^4.
\end{align}

\noindent In both gas and radiation pressure dominated cases, the factor has a typical value of $10^{-2}$, suggesting that SMDs are likely to contract significantly once rotational support is completely lost. A more accurate result can be found by solving Eq.~\ref{eq:energy_contraction} numerically.

\begin{figure}
    \centering
    \includegraphics[scale=0.7]{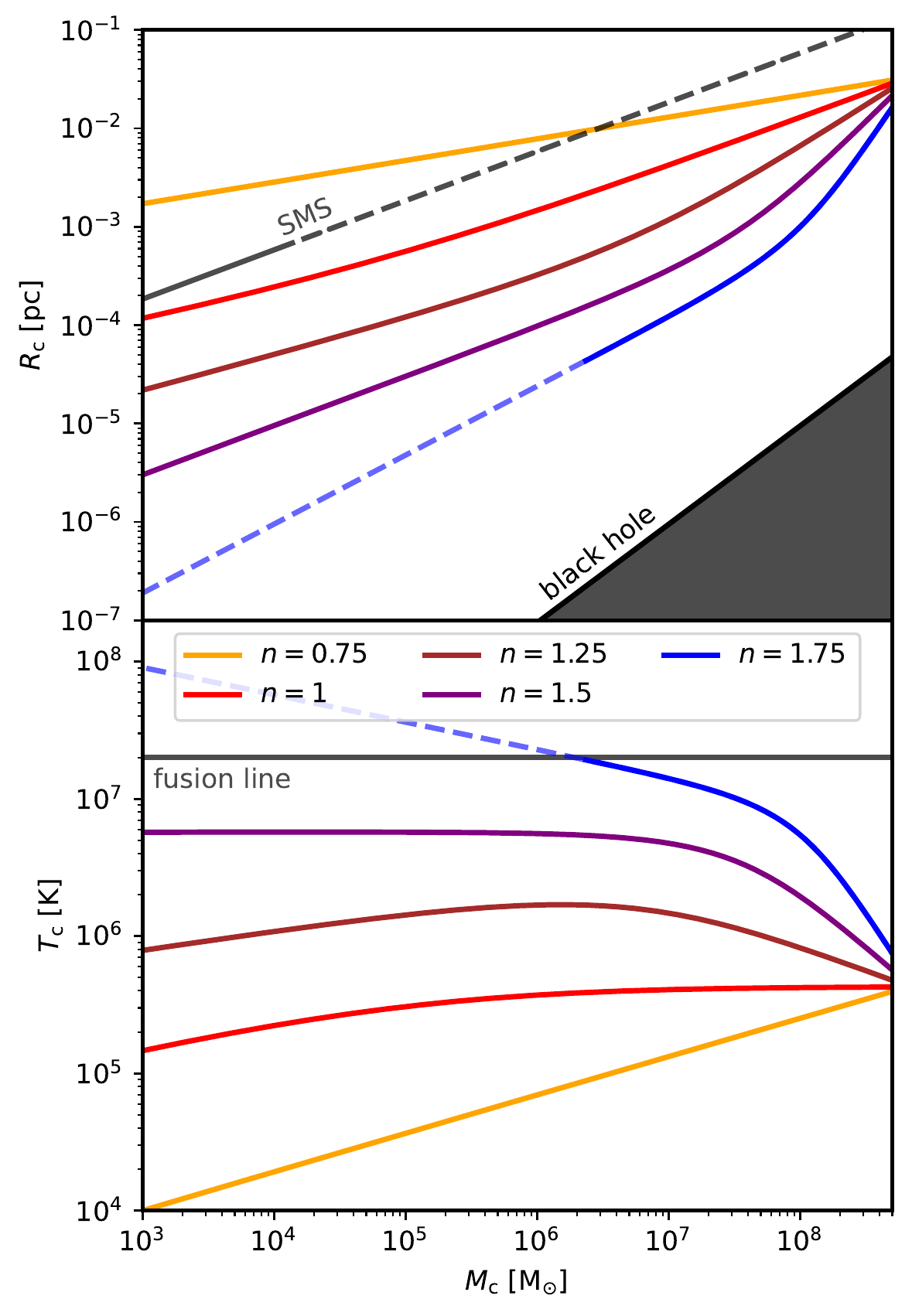}
    \caption{The mass-radius and temperature relations for the core of an SMD with mass  $10^9$ M$_{\sun}$ and size 1 pc. The coloured lines represent realisations of the disc with different enclosed mass profiles (Eq.~\ref{eq:Mencl}). As more shells are accreted, the core increases in size, mass and internal energy. In the upper panel, the grey line represents the expected mass radius relation for a supermassive star, extrapolated beyond its known validity range by the dashed line \citep[see, e.g.][]{2012hosokawa}. The boundary of the grey triangle denotes the mass-radius relation of a BH. In the lower panel, the grey line denotes the ignition temperature for hydrogen fusion, which is approximately $2\times 10^7$ K.}
    \label{fig:core_radius}
\end{figure}

Once a given shell has contracted to its new equilibrium radius, it has by construction added exactly the amount of mass and thermal energy needed for a core of mass $M_0$ to be in hydrostatic equilibrium, where $M_0$ is the enclosed mass within the shell's initial radius $r_0$. This means that, with the aid of the solutions to Eq.~\ref{eq:energy_contraction}, we can find a simple relationship for the mass ($M_{\rm c}$) and size ($R_{\rm c}$) of the hydrostatic core once a given shell is added to it:

\begin{align}
    M_{\rm c}(R_{\rm c}) = M_{\rm d} \left(\frac{r_0(R_{\rm c})}{R_{\rm d}} \right)^{3-n},
\end{align}

\noindent where $r_0$ is the initial radius of the specific shell that achieves equilibrium at the radius $R_{\rm c}$. We can find explicit solutions for the gas and the radiation dominated regimes by simply inverting Eqs.~\ref{eq:contraction_gas} and~\ref{eq:contraction_rad}:

\begin{align}
    M_{\rm c}^{\rm{gas}}(R_{\rm c}) &= M_{\rm d}\left(\frac{R_{\rm c} }{ \mathcal{K}_{\rm{gas}} R_{\rm d}} \right)^{6/n -2}, \\
    M_{\rm c}^{\rm{rad}}(R_{\rm c}) &= M_{\rm d}\left(\frac{R_{\rm c} }{ \mathcal{K}_{\rm{rad}} R_{\rm d}} \right)^{(n-3)/(2n -1)}.
\end{align}
As expected, the core's mass approaches the mass of the whole SMD once its radius becomes $\mathcal{K}R_{\rm d}$. Numerical results for the mass-radius relation that include both gas and radiation pressure are shown in the top panel of Figure~\ref{fig:core_radius}, for several choices of the parameter $n$. In general, the initial degree of mass concentration strongly determines the growth of the cores. At high masses however, the cores tend to approach the same radius regardless of the value of $n$. This behaviour is due to the cores always transitioning from being gas pressure to radiation pressure dominated. Therefore, their radius at a mass of $10^9$ M$_{\sun}$ must approach the value $\mathcal{K}_{\rm{rad}}R_{\rm d}$, which is derived in the same limit. The total amount of mass and internal energy is always dominated by the accretion of new layers (for $n \lsim 2$). Therefore, the temperature within the core must be mostly determined by the temperature of the last accreted shell, which is by construction equal to the virial temperature of the enclosed mass. We can therefore assume that the cores always re-establish local isothermal equilibrium at the virial temperature appropriate for their new radius $r_{\rm{eq}}$, every time a new shell is added.

As discussed previously, the mass-radius relationship for the core relies exclusively on the assumptions of adiabaticity and virialisation of rotational energy. It is therefore a generic feature of the evolution of SMDs that does not depend on the details of angular momentum loss. Adiabaticity can be broken in the presence of significant amounts of nuclear fusion, which would provide an additional heating term. For the purposes of this paper, we adopt a temperature limit of $\sim 2\times 10^7$ K for the ignition of hydrogen within the core \citep[see, e.g.][]{2004palla}, and consider any core whose virial temperature exceeds this limit to be affected by nuclear fusion. As is visible in the lower panel of Figure~\ref{fig:core_radius}, only SMDs with high degrees of mass concentration ($n \gsim 1.7$) produce cores in which nuclear fusion is likely. In this case, the cores would tend to heat up and increase in radius faster than what our relations predict, possibly evolving along supermassive star tracks \citep[see, e.g.][]{2012hosokawa,2021lionel}. We refer the reader to section~\ref{sec:caveats} for a more thorough discussion.

\subsection{Formation of a BH via the GR instability}\label{sec:AM_collapse}

As more and more shells lose angular momentum, the hydrostatic core increases in size, mass and internal energy. The relative contribution between gas pressure and radiation pressure changes in favour of the latter. As discussed in Section~\ref{sec:TB:GRinst}, the criterion for GR instability is met whenever the adiabatic index is sufficiently close to the critical value of 4/3. For the purposes of this work, we deem the core to be GR-unstable whenever its average adiabatic $\gamma_{\rm c}$ index crosses over a phenomenological value $\gamma_{p}$ given by the following equation:

\begin{align}
    \label{eq:critadiabatic}
    \gamma_{\rm{p}} = 4/3 + \frac{1}{2}\frac{2 G M_{\rm c}}{c^2 R_{\rm c}},
\end{align}

\noindent where the factor 1/2 approximates the exact value of 19/42 appropriate for homogeneous density distributions, i.e the average density of the core\footnote{Changing the prefactor $\kappa$ does not affect the results, since even in the case of a polytropic index of 3, $\kappa$ is at most $\sim 1.1$ \citep[][]{1964chandra}.}. For the purposes of this analysis we assume that, once the core becomes GR-unstable, it will collapse retaining most of its mass and producing a massive BH. Since it is dynamical, the collapse must occur on a time-scale comparable to the local free fall time. In order to find the mass at which the cores in SMDs become unstable, we compute an average adiabatic index by using the mass-radius relation, the virial temperature and Eq.\ref{eq:adiabaticindex}. We then solve Eq.~\ref{eq:critadiabatic}, evaluated at the appropriate mass, radius and average adiabatic index that is associated to the core.

\begin{figure}
    \centering
    \includegraphics[scale=0.7]{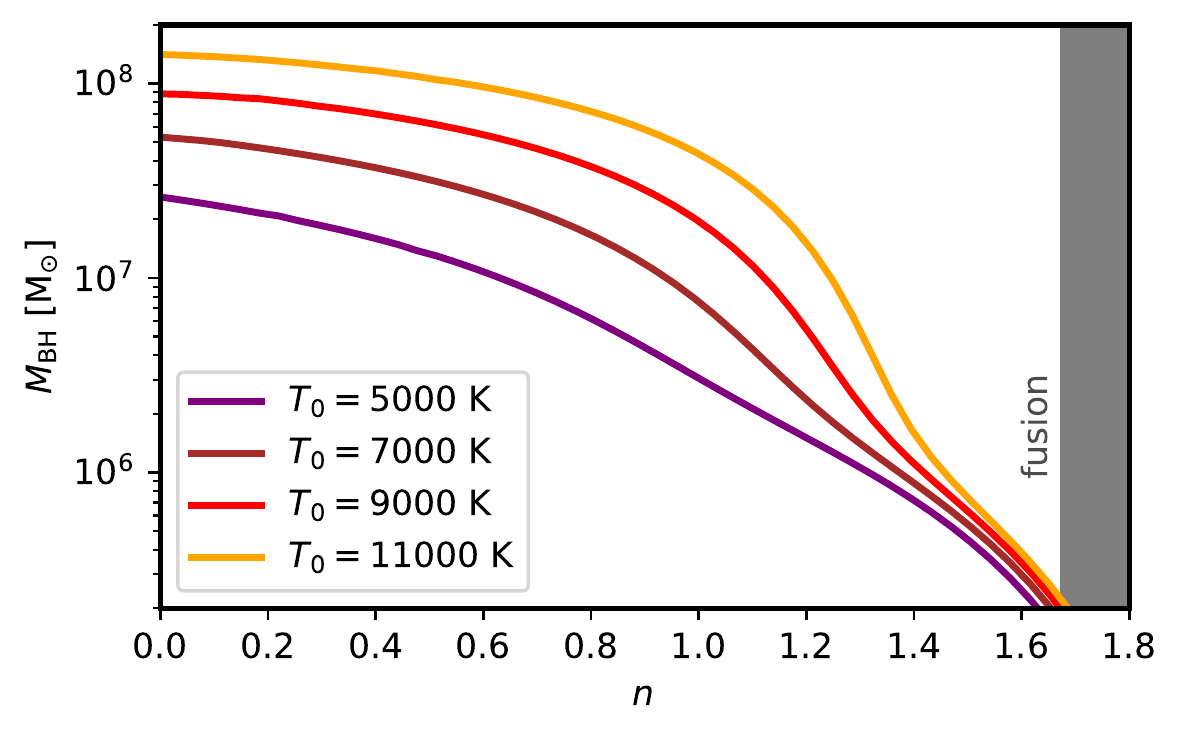}
    \caption{The mass of the final BH seed formed in an SMD with mass  $10^9$ M$_{\sun}$ and size 1 pc. The coloured lines represent realisations of the disc with different initial temperatures, plotted as a function of the mass profile parameter $n$ (Eq.~\ref{eq:Mencl}). More concentrated SMDs produce smaller BH seeds, while shallower discs produce larger ones. The seeds are formed rapidly, at most $\sim 10^5$~yr after the formation of the SMD at $z \sim 10$. These exceptionally heavy DC BHs are not only consistent with, but can easily explain even the largest observed high-redshift quasars.}
    \label{fig:BH_mass}
\end{figure}

The results of our computations are visualised in figure~\ref{fig:BH_mass}.  Typically, the GR-unstable cores collapse from an initial compactness of $\sim 100$, far from the strong gravity regime of GR. The final BH seed masses strongly depend on the value of the parameter $n$ and can range from $\sim 10^6$ M$_{\sun}$ to $10^8$ M$_{\sun}$. Within this range is the direct formation of exceptionally heavy seeds, all the way up to \textit{fully fledged SMBHs}. The model is therefore not only consistent with the largest observed high red-shift quasars, but can easily explain their masses, since it only requires the seeds to accrete at Eddington levels for a few Salpeter time-scales rather than dozens. The additional mass can be provided by the rest of the SMD, which must eventually settle into an accretion disc around the newly formed seed, likely powering quasar emission.

In general, more concentrated SMDs produce smaller BH seeds, while SMDs with shallow density profiles produce large BH seeds. The trend is simply due to the fact that the GR instability is triggered at smaller scales if more mass is concentrated towards the centre of the disc. The initial temperature of the SMD has a noticeable effect on the final BH mass in discs with shallower density profiles ($n \lsim 1.4$). This is a consequence of the relative amount of mass in the core that is primarily supported by gas pressure versus radiation pressure. A higher initial temperature produces less dense, puffier cores that become unstable at larger scales, thus enclosing more mass. The total initial mass and radius of the disc have a negligible effect on the final masses, since the general density scale of SMDs is set by the transition between the optically thin and adiabatic evolution regimes (see Section~\ref{sec:TB_supermassive_disc}) and its value cannot deviate significantly from $\sim 10^{-11}$ Kg/m$^3$. Less dense discs with a size of $\sim 1$ pc might still produce some form of compact objects, just not via the specific mechanism proposed in this paper. Much denser disc are unlikely to exist at $z \sim 10$ due to the limitations of dark matter halo masses. If one still allowed the density to vary, denser SMDs would typically produce slightly lighter BHs, while the reverse is also true. We stress that the collapse of these exceptionally heavy seeds is \textit{rapid}. From the formation of the SMD at $z \sim 10$, at most $\sim 5\times 10^5$~yr are required for the core to approach the GR-unstable regime. The core then collapses on a local dynamical time, which ranges from days to months for the relevant densities. To summarise all the possibilities, we provide a simple formula which accurately fits (to a factor of order unity) the numerical results within a range of reasonable parameters for the initial SMD, i.e $2<n<0$ and $5000$ K $< T_0<11000$ K. The final BH masses ($M_{\rm{BH}}$) are given by the following expression:

\begin{align}
    \label{eq:fit}
    \frac{M_{\rm{BH}}}{\rm{M}_{\sun}} &\sim 3 \times 10^7 f(T_0,n)  \\
   f(T_0,n) &=  \left(1 - \frac{n}{2}\right)^{2.8} \nonumber\\ &+ \frac{2}{3}\left(\frac{T_0}{7000 \, \rm{K}}\right)^{4}  \left(1 - \left(\frac{4 n}{5}\right)^2 \right)^{1.8},
\end{align}

\noindent where the latter term must be set to 0 as soon as it becomes undefined ($n> 1.25$). We expect this fitting formula to be accurate (within the limits of our analysis) for SMDs in which the core is unlikely to ignite, i.e. with a parameter $n \lsim 1.7$. It can be applied as a sub-grid model in cosmological simulations of major mergers as well as population models that have the aim of reproducing the observed statistics of high-redshift quasars and the BH mass function \citep[see, e.g.][]{2014bonoli,2023Spinoso}.

\section{Discussion}\label{sec:discussion}

\subsection{Smoking gun observational signatures}\label{sec:observation}

The hydrostatic cores described in Section~\ref{sec:AM:accretion} grow in size over the course of the accretion process, reaching radii of $\sim 10^{-5}$ pc to $10^{-2}$ pc before collapsing via the GR instability. Because of their large size and relatively high temperature ($\sim 10^6$ K), we expect them to be bright sources of thermal radiation. Neglecting for a moment the fact that the cores are embedded in opaque discs, we can describe their bolometric luminosity, $L_{\rm c}$, by means of the Stefan-Boltzmann law:

\begin{align}
    \label{eq:BBlum}
    L_{\rm c} = 4 \pi R_{\rm c}^2 \sigma_{\rm B} T_{\rm c}^4.
\end{align}

\noindent The growth of the cores occurs on a time-scale comparable to the angular momentum loss time-scale, i.e. $\lsim 10^5$~yr. Over this period, the temperature variation is limited (as shown in the bottom panel of figure~\ref{fig:core_radius}) and we expect the luminosity to roughly increase as $\sim R_{\rm c}^2$. Once the GR instability is reached, the core collapses on a much shorter dynamical time-scale. This rapid contraction causes a violent increase in temperature, both due to adiabatic heating and virialisation by turbulent dissipation. The temperature in a contracting adiabatic system scales as $T \sim R^{-1}$. We therefore expect the luminosity of the cores to roughly scale as $R_{\rm c}^{-2}$ during the collapse, mirroring the trend during the growth phase. However, a couple of effects can taper down the luminosity in the latter stages of collapse. Firstly, the rapid change in size will detach the core from the rest of the disc. It will then be able to freely radiate thermal energy into the cavity it leaves behind. This introduces a cooling term in the energy balance, which is described by Eq.~\ref{eq:BBlum}. Secondly, whenever temperatures in the core reach values of $\sim 10^9$ K, pair annihilation and the URCA process can induce a large outward flux of neutrinos \citep[see, e.g.][]{1989Itoh,1996Qian,2004Dutta}. Neutrinos can carry a significant amount of energy, adding an additional cooling term that is highly sensitive to temperature. For the purposes of this work, we adopt a simple approximation to the neutrino cooling rate $Q_{\rm n}$, also used in \cite{2006begelman} in the context of DCMs:

\begin{align}
    Q_{\rm n} \approx 3\times 10^{15} \left(\frac{T_{\rm c}}{10^9 \, \rm{K}} \right)^{9} \, \left[\frac{\rm{erg}}{\rm{s} \, \rm{cm}^3 }\right].
\end{align}

\begin{figure}
    \centering
    \includegraphics[scale=0.7]{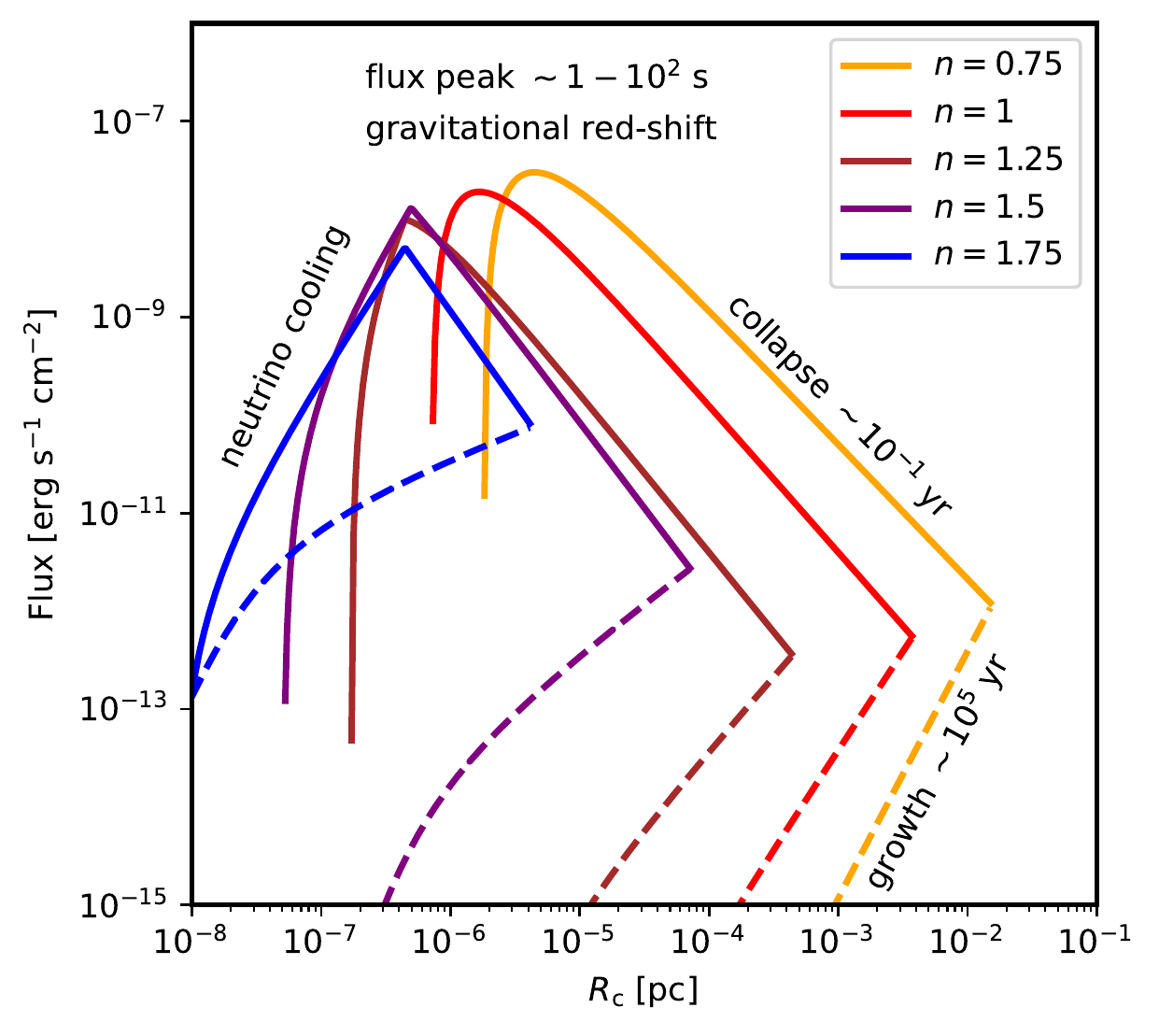}
    \caption{The bolometric flux tracks of SMD cores observed at $z = 10$, plotted as a function of the core radii for different values of the parameter $n$. The growth phase (dashed lines) ends once the GR instability is reached and dynamical collapse is triggered. The luminosity increases dramatically during the collapse, until it is suppressed by gravitational redshift or neutrino cooling (solid lines). The resulting x-ray and neutrino transients would constitute a smoking gun signature of the merger-driven DCM. The tracks assume that the radiation from the cores is able to escape the host proto-galaxy without being absorbed or scattered.}
    \label{fig:em_signature}
\end{figure}

\noindent Finally, gravitational redshift, $z_{\rm g}$, can significantly dim the luminosity of a system that is approaching its own gravitational radius. The value for gravitational redshift is given by the following formula:

\begin{align}
    z_{\rm g} = \left(1 - \frac{r_{\rm S}}{R_{\rm c}} \right)^{-1/2} -1,
\end{align}

\noindent and it decreases the bolometric luminosity by a factor $(1 + z_{\rm g})^4$ \citep[see, e.g.][]{1993peebles}. In Figure~\ref{fig:em_signature} we show the bolometric luminosity tracks of several SMD cores as a function of their radius, for different values of the parameter $n$. To produce the tracks, we consider the effects of accretion, adiabatic heating, virialisation, black body radiation, neutrino cooling and gravitational redshift on the thermodynamics of the cores. We express the results in terms of the expected flux at a luminosity distance appropriate for a $z = 10$ source. This would represent the flux as seen by an observer on the Earth, \textit{assuming} that radiation is able to freely escape the host proto-galaxy. We consider this process to be a first step in assessing the direct detectability of a merger-driven DC event as a source of radiation, though significantly more modelling is required for a realistic prediction.

As can be seen in the figure, the tracks initially follow the growth of the core. During this phase, the flux typically peaks at values of $\sim 10^{-12}$ ergs s$^{-1}$ cm$^{-2}$ and is mostly concentrated in the ultraviolet and soft X-ray bands. The duration of this phase depends on the angular momentum loss mechanism, which can range from $\sim 10^2$~yr to $\sim 10^5$~yr.
Once the GR instability is triggered, the cores start to contract and the flux rapidly increases by several orders of magnitudes due to the rise in temperature. The signatures of this phase are extremely bright, hard x-ray (and potentially gamma-ray) transients that peak at a bolometric flux of $\sim 10^{-8}$ ergs s$^{-1}$ cm$^{-2}$. The duration of this phase depends on the size of the core as it becomes unstable. In general, the peak luminosity is limited to an extremely short period of  a few seconds to minutes. During this phase, the core detaches from the rest of the disc due to its rapid collapse, leaving behind a cavity at the centre of the disc. Depending on the configuration and the disc geometry, it possible that radiation may escape from the proto-galaxy, as suggested by the cartoon in Figure~\ref{fig:cartoon}. The very last stages of the tracks differ qualitatively for SMDs with shallow ($n\lsim 1$) or steep ($n\gsim 1$) initial density profiles. For the former, the flux is smoothly reduced by gravitational redshift in the moments before a BH is formed. For the latter, the temperatures are high enough for neutrino cooling to suddenly become efficient. While the electromagnetic luminosity is reduced, neutrino cooling induces a flux of particles with a scale of $\sim 10^{-7}$ ergs s$^{-1}$ cm$^{-2}$. Neutrinos are much more likely to freely exit the proto-galaxy, and the collapse of an SMD core could in principle be detectable as a point source transient by current and future instruments \citep[see, e.g.][]{2011ice,2013ice,2021ice2}.

The electromagnetic and particle transients are not the only smoking guns of the merger-driven DC scenario. If a small amount of angular momentum is retained within the core at the moment it becomes unstable (which we discuss in Section~\ref{sec:caveats}), we would expect its collapse to not be exactly spherically symmetric, i.e. have a non vanishing quadrupole mass moment. According to GR, this would imply the generation of a gravitational wave burst \citep[][]{Einstein2,Einstein1,1973grav,1977chia,2011fryer}. The gravitational collapse of oblate spheroids has been studied in several works \citep[see, e.g.][]{1965lin,1968fujimoto,1972larson,1983goodman}. For the purposes of this analysis, we assume that cores contract homologously, i.e. the ratio between their semi-major axes, $a$, and semi-minor axes, $b$, remains constant. The quadrupole mass moment, $\mathcal{M}$, of an oblate spheroid is given by the following formula:

\begin{align}
    \mathcal{M} &= \frac{2}{5}M_{\rm c} R_{\rm c}^2 \epsilon^2 \\
    \epsilon^2 &= \frac{1 - (a/b)^2}{(a/b)^{4/3}}
\end{align}

\noindent where we replaced the geometric-mean radius of the spheroid with the core radius $R_{\rm c}$ and defined an ellipticity $\epsilon$. In the quadrupole approximation, the typical amplitude, $h$, of a gravitational wave depends on the second derivative of the mass quadrupole:

\begin{align}
    \label{eq:gwampli}
     h &\sim \frac{G}{c^4 D_{\rm l}} \Ddot{\mathcal{M}},
\end{align}

\noindent where $D_{\rm l}$ is the luminosity distance. Dimensionally, the amplitude produced by a collapsing spheroid is therefore given by the following simple expression:

\begin{align}
    h &\sim \epsilon^2\frac{G M_{\rm c}}{c^2 D_{\rm l}} \left(\frac{v}{c}\right)^2 \\
    &\approx 4 \times 10^{-21} \left(\frac{\epsilon}{10^{-2}} \right)^2\left(\frac{
    M_{\rm c}}{10^7\, \rm{M}_{\sun}} \right)\left(\frac{10\, \rm{Gpc}}{D_{\rm l}} \right) \left(\frac{v}{c} \right)^2,
\end{align}

\noindent where we introduced a characteristic velocity $v \sim \Dot{R}_{\rm c}$, which becomes comparable to the speed of light during the late stages of collapse. Scaled with values typical to cores in SMDs, we find that the amplitude of the burst is comparable to the sensitivity of both existing ground based and future space based detectors for an ellipticity of $\epsilon \gsim 10^{-2}$. The typical frequency of a burst, $f_{\rm b}$, at its loudest point (when $v \sim c$) is comparable to the reciprocal of the light-crossing time for the final BH. For values typical to SMD cores it reads:

\begin{align}
    f_{\rm b} \sim \frac{c^3}{2 G M_{\rm c}} \approx 10^{-2} \left(\frac{10^7\, \rm{M}_{\sun}}{
    M_{\rm c}} \right)\, \left[\rm{Hz} \right],
\end{align}

\noindent which places the bursts squarely in the preferred frequency band for the upcoming Laser Interferometer Space Antenna \citep[][]{lisa2017,2019baker,2022lisa,2022auclair} and TianQin detectors \citep[][]{2016tianqin,2021tianqin}.

We can roughly estimate the event rate at which the transients discussed above are expected to occur. An estimate of the rate of appropriate major mergers that can produce SMDs is provided in \cite{2019Mayer}, ranging between 10$^{-7}$ to 10$^{-5}$ mergers per Mpc$^3$. Here we consider an interesting redshift window between $z = 7$ and $z = 12$, where the lower value is given by the furthest observed high mass quasars while the larger one is an arbitrary boundary above which major mergers are not expected. We can simply multiply the merger rate with the co-moving volume between $z=7$ and $z=12$, which amounts to $\sim 1500$ Gpc$^3$ for a flat $\Lambda$CDM universe \citep[][]{2006wright}. We find that the total amount of appropriate mergers amounts to $\sim 10^5$ to $\sim 10^7$. Dividing the number by $500$ million years, i.e. the elapsed time between $z=7$ and $z=12$, we find that we should expect a rate of $\sim 10^{-2}$ to $\sim 1$ gravitational, neutrino and electromagnetic transients per year, assuming that every suitable major merger inevitably leads to a DC event\footnote{As shown in \cite{2014bonoli}, these rates are consistent with the statistics of high red-shift quasars.}. The likelihood for detecting a counterpart is boosted significantly with instruments sensitive enough to observe the growth and the initial collapse phases of the cores due to their longer duration. \newline \newline  We look forward to the feed-back of the x-ray, neutrino and gravitational wave astronomy communities to assess whether these signatures are detectable in the foreseeable future, or might even be present in existing wide field surveys.

\subsection{Caveats}\label{sec:caveats}

The largest caveat of our analysis is the simplistic treatment of the hydrostatic cores, which are likely to settle into structures resembling SMSs.
The arguments used in section \ref{sec:AM:accretion} only allow for an order of magnitude estimate of their properties, e.g. the radius, mass and energy content.
A more precise analysis would require full solutions to the equation of hydrostatic equilibrium, which necessarily implies the use of numerical models. In particular, the latter show that the structure of SMSs depend critically on their mass accretion rate. For rates greater than $0.01$~M$_\odot$~yr$^{-1}$, accretion prevents thermal relaxation on fully convective, polytropic structures. The SMS evolves instead as a 'red supergiant protostar' \citep{2012hosokawa,hosokawa2013,haemmerle2018a}, whose internal structure consists of a convective core produced by H-burning, a radiative zone containing most of the stellar mass, and a convective envelope that covers a dominant fraction of the photospheric radius.
These structures can be approximated by 'hylotropes' \citep{begelman2010}, in particular for rates $\gtrsim10$ M$_\odot$ yr$^{-1}$, at which the contraction is adiabatic \citep{haemmerle2019c}. The hydrodynamical simulations of \cite{mayer2010,Mayer2015} show accretion rates up to $\sim10^5$ M$_\odot$ yr$^{-1}$ down to a fraction of a pc during the assembly of the SMD. Gravitational free-fall can maintain such rates down to the stellar surface only if its mass exceeds $\sim10^5$ M$_\odot$. Milder rates up to $\sim1000$ M$_\odot$ yr$^{-1}$ can be driven if the attractor's mass exceeds $\sim1000$ M$_\odot$ \citep{lionel2021}, which is comparable to our estimates for the mass of the initial hydrostatic cores in SMDs. A dynamical angular momentum loss mechanism (as discussed in section~\ref{sec:TB}) could supply accretion rates close to the maximum value allowed by gravitational free-fall. At such rates, masses up to $10^8-10^9$ M$_\odot$ could be reached before the final collapse, which corresponds to similar estimates as those of section \ref{sec:AM_collapse}.

Another important factor we neglect to model explicitly is rotation. Because of the prominent role of radiation pressure, SMSs must have surface velocities lesser than $10\%$ of the Keplerian velocity \citep{haemmerle2018b,haemmerle2019a}.
As a consequence, the centrifugal acceleration represents at maximum a percent of the gravitational acceleration, implying that rotation has a negligible impact on the hydrostatic structure and that deformation remains small. On the other hand, rotation impacts significantly the dynamical stability of the structure \citep{fowler1966,bisnovatyi1967,baumgarte1999b,shibata2016a,2021lionel}. The mass at which rapidly accreting SMSs reach the GR instability depends on both the accretion rate of mass and angular momentum. Without any angular momentum, reaching masses $\gtrsim10^6$ M$_\odot$ before becoming unstable typically requires accretion rates larger than $1000$ M$_\odot$ yr$^{-1}$ \citep{2020lionel}. However, even a small amount of angular momentum postpones this limit by orders of magnitude, and maximally rotating SMSs can remain stable potentially up to $\sim10^9$ M$_\odot$ \citep{fowler1966,2021lionel}. Since the growth of the cores described in section \ref{sec:AM:accretion} results from the accretion of a rotationally supported disc, it is likely that some amount of angular momentum will be retained even in the final hydrostatic structure. Some combination of high accretion rates and retention of angular momentum is exceedingly likely in SMD cores, strongly supporting that the high final BH masses reported in Figure~\ref{fig:BH_mass} are realistic estimates despite the simplifying assumptions.

The final issue that requires further discussion is fragmentation. As shown in MI and MII, the innermost region of SMDs (whithin $\sim 1$ pc) is stabilised by the extreme shear forces, preventing the formation of stars. However, it is possible that significant fragmentation may occur at the outer edges, where the Toomre parameter approaches its critical value. Thus, the consequence of fragmentation would be to enable the formation of a stellar disc with a typical size of a few pc. Understanding the interplay between fragmentation, migration and relaxation processes within the SMD and its stellar counterpart would require significant modelling, and would be best tackled by a purpose built numerical simulation. If the scenario lead to a significant flux of stars towards the central region, it could perhaps enable an extreme version of the stellar bombardment model proposed in \cite{2020tagawa}, in which runaway mergers of stars take over the mass and energy balance of the growing SMD core. Both the dynamics of the stellar disc and the evolution and structure of SMD cores under bombardment are extremely interesting and deserve further investigation.

\section{Summary \& Conclusion}\label{sec:conclusion}

In this work we revisited the merger-driven scenario in order to asses its promise as a DCM. In Section \ref{sec:TB}, we show that the thermodynamical evolution of SMDs is fundamentally different than more typical proto-galactic discs. Firstly, SMDs are optically thick and therefore evolve adiabatically over periods comparable with the diffusion time-scale of $\lsim 5\times 10^5$~yr (see Eq.~\ref{eq:difftimes}). Fragmentation is therefore naturally suppressed without the requirement of an external source of ionising radiation. Secondly, SMDs are expected to lose angular momentum on time-scales comparable or shorter than a diffusion time-scale, both in the case of dynamical as well as secular mechanisms (see Eq.~\ref{eq:diffvisc} for the case of turbulent viscous dissipation). In combination with the adiabaticity, this implies that our analysis is independent of the choice of an angular momentum loss mechanism. We sumarise the DCM as follows: As an SMD loses rotational support, more and more concentric shells accrete onto to a central spherical hydrostatic core (see Section~ \ref{sec:AM:accretion}). Every shell adds a specific amount of mass and internal energy, which determines the core's mass-radius and temperature relation (shown in figure~\ref{fig:core_radius}). Eventually, the cores approach the GR instability regime and collapse into a compact object, producing several observational counterparts that can be considered smoking gun signatures of the model.
The main findings of our analysis are listed here:

\begin{itemize}

    \item The mass of BHs produced in the merger-driven scenario has a typical scale of  a few $10^7$ M$_{\sun}$, ranging from $10^6$ M$_{\sun}$ to $10^8$ M$_{\sun}$ depending on the initial disc configuration (see figure~\ref{fig:BH_mass}). SMDs are therefore optimal locations to produce fully fledged SMBHs by direct collapse. The collapse occurs very rapidly, at most $\sim 5\times 10^5$~yr after the formation of an SMD at $z \sim 10$. A simple fitting formula for the final BH mass is provided by Eq.~\ref{eq:fit} and can be implemented as a sub-grid model in future simulations.
    
    \item The collapse of an SMD core produces a characteristic luminosity curve, shown in figure~\ref{eq:BBlum}. It exhibits an extremely bright ($\sim 10^{-8}$ erg s$^{-1}$ cm$^{-2}$) but short ($\sim $ few s) x-ray transient. In certain cases, the electromagnetic signature might be accompanied by a burst of neutrinos ($\sim 10^{-7}$ erg s$^{-1}$ cm$^{-2}$ in flux) and of gravitational wave radiation ($\sim 10^{-21}$ in amplitude, see Eq.~\ref{eq:gwampli}), potentially detectable by current and future instruments. If identified, these counterparts would represent a smoking gun signature of the merger driven DCM. Their expected rate is $\sim 10^{-2}$ to $\sim 1$ per year.
    
\end{itemize}

The merger-driven scenario has the advantage of not requiring any specific assumptions on the thermodynamics or on the angular momentum loss mechanisms of proto-galactic discs, provided that they are massive enough to behave adiabatically over long time-scales. This is a unique feature, which is at odds with other DCMs. In this paper, we have shown that once an SMD forms, it is inevitable that it will produce a large BH on a very rapid time-scale, without requiring any further additional assumption other than  conventional hydrodynamics
of ideal gas flow and the large-scale dynamics imposed by the
merger. The trade-off is that only relatively rare major mergers
of galaxies well exceeding the typical mass of star forming galaxies
at $z > 10$ can
form SMDs, implying  that this flavour of DC must be rare by construction 
\citep[see][for an estimate of the occurence rate  of such systems, and its agreement with the abundance of high-z quasars.]{2014bonoli,2019Mayer}.

Returning to the question posed in the introduction, the merger-driven scenario implies the possibility that high red-shift BHs with varying masses might form by different mechanisms. In this scenario, an uncountable number of smaller BH seeds can form early, by any of the several models present throughout the literature, including the other more conventional DC channels.
These might then slowly grow by accretion or by repeated mergers to become one of the many BHs, ranging from $10^5$ M$_{\sun}$ to 10$^{8}$ M$_{\sun}$, that we identify at the centre of galaxies in our local universe. The BH mass function at high redshift could therefore start off rather ``under-weight'', and remain so up until the sudden appearance of merger-driven DC BHs at $z \sim 10$. These would form fast and late, by a completely separate mechanism from the majority. They would instantly surpass their older, tamer counterparts, becoming visible as $10^9$ M$_{\sun}$ quasars at $z \sim 7$ and possibly explaining the formation of the exceptionally massive quasars such as TON618, Holmberg 15A and many others.
Incidentally, these heaviest of SMBHs tend to reside in elliptical early type galaxies, the likely outcome of the massive mergers required to form SMDs in the first place \citep[see, e.g.][]{1970ton,2003cattaneo,2013mcconnell,2014a15,2019yoon}. \newline \newline

In conclusion, it is hard to argue which answer would be more satisfying: a single DCM, in which the most massive outliers are explained as improbable statistical flukes? Or rather the co-existence of several models, which can explain BHs in different mass ranges more naturally? Perhaps the answer must come from astronomical observation. With a new generation of detectors coming, we are certainly hopeful that a careful analysis of the redshift-dependent BH mass function, as well as the constant monitoring of the electromagnetic, neutrino and gravitational wave skies will be sufficient to untangle the question of early BH formation, be it merger-driven or otherwise.

\section*{Acknowledgements}
The authors acknowledge support from the Swiss National Science Foundation under the Grant 200020\_192092. We acknowledge Pedro R. Capelo, Andrea Derdzinski, and Mudit Garg for insightful discussions.
LH has received funding from the European Research Council (ERC) under the European Union's Horizon 2020 research and innovation programme
(grant agreement No 833925, project STAREX).
RSK acknowledges financial support from the European Research Council in the ERC Synergy Grant `ECOGAL' (project ID 855130), from the Heidelberg Cluster of Excellence `STRUCTURES' (EXC 2181 - 390900948), from the German Science Foundation (DFG) via the Collaborative Research Center `The Milky Way System’ (SFB 881, Funding-ID 138713538, subprojects A1, B1, B2, B8),  and from the German Ministry for Economic Affairs and Climate Action in the project `MAINN' (funding ID 50OO2206). We thank the
Kavli Institute for Theoretical Physics at UC Santa Barbara for
hospitality during the development of this work,in particular the program
BINARY22 supported in part by the National Science Foundation under Grant No. NSF PHY-1748958.

\section*{Data availability}
Our python scripts will be shared upon reasonable request.

\scalefont{0.94}
\setlength{\bibhang}{1.6em}
\setlength\labelwidth{0.0em}
\bibliographystyle{mnras}
\bibliography{main}
\normalsize


\bsp 
\label{lastpage}
\end{document}